\global\def\draftcontrol{0}

   \def\versionno{ baryonic branch inflation -- draft   }

\catcode`\@=11

\expandafter\ifx\csname draftcontrol\endcsname\relax\global\def\draftcontrol{0}
\fi

{\count255=\time\divide\count255 by 60
\xdef\hourmin{\number\count255}
\multiply\count255 by-60\advance\count255 by\time
\xdef\hourmin{\hourmin:\ifnum\count255<10 0\fi\the\count255}}
\def\draftdate{\number\month/\number\day/\number\year\ \ \ \hourmin }

\newcommand\makepapertitle{\par
  \begingroup
    \renewcommand\thefootnote{\@fnsymbol\c@footnote}%
    \def\@makefnmark{\rlap{\@textsuperscript{\normalfont\@thefnmark}}}%
    \long\def\@makefntext##1{\parindent 1em\noindent
            \hb@xt@1.8em{%
                \hss\@textsuperscript{\normalfont\@thefnmark}}##1}%
     \newpage
     \global\@topnum\z@   
     \@makepapertitle
     \thispagestyle{empty}\@thanks
  \endgroup
  \setcounter{footnote}{0}%
  \global\let\thanks\relax
  \global\let\makepapertitle\relax
  \global\let\@makepapertitle\relax
  \global\let\@thanks\@empty
  \global\let\@author\@empty
  \global\let\@date\@empty
  \global\let\@title\@empty
  \global\let\title\relax
  \global\let\author\relax
  \global\let\date\relax
  \global\let\and\relax
  \def\version{\let\version\@version\@gobble}
}
\def\@makepapertitle{%
  \newpage
   \ifnum\draftcontrol=1 {}
   \version\versionno
   \vskip 3em%
   \else
   \hfill\hbox to 3cm {\parbox{4cm}{\@pubnum}\hss}%
   \vskip 3em%
   \fi
   \begin{center}%
   \let \footnote \thanks
     {\LARGE {\@title}}%
     \vskip 1.5em%
     {\normalsize
       \lineskip .5em%
       \begin{tabular}[t]{c}%
         \@author
       \end{tabular}\par}%
     \vskip 1.5em%
     {\@bstract}%
     \end{center}%
     \vskip 1.5em
     \@date%
   \par
}

\gdef\@pubnum{}
\def\pubnum#1{%
  \gdef\@pubnum{#1}}

\gdef\@bstract{}
\def\Abstract#1{%
  \gdef\@bstract{%
   \parbox{\textwidth-0pc}{%
   \centerline{\bf Abstract}\penalty1000%
\kern.2cm%
\noindent
\renewcommand\baselinestretch{1.0}%
{#1}}}
}

\def\ps@paper{\let\@mkboth\@gobbletwo%
     \ifnum\draftcontrol=1
	\def\@oddfoot{\hbox to \textwidth{\tiny \versionno \hfil\tiny\draftdate}%
	\hskip -\textwidth \hbox to \textwidth{\hfil\rm\thepage\hfil}}%
     \else\def\@oddfoot{\hbox to \textwidth{\hfil\rm\thepage\hfil}}
     \fi
     \let\@evenfoot\@oddfoot
}

\def\body{\clearpage
          \pagestyle{paper}
	}

\def\@version#1{\ifnum\draftcontrol=1
\typeout{}\typeout{#1}\typeout{}
\vskip3mm\centerline{\hbox{\fbox{\normalsize{\tt DRAFT -- #1 -- }
                   {\draftdate}}}}\vskip3mm
\fi}
\let\version\@version
\long\def\eqlabel#1{\ifnum\draftcontrol=1
                    \tag@false  
                    \tag*{(\theequation) \hbox to -0.2cm{\hspace{0cm}\small{#1}\hss}}
                    \refstepcounter{equation}
                    \edef\@currentlabel{\theequation}
                    \ltx@label{#1}          
                    \else
                    \label{#1}
                    \fi
                    }
\let\st@bibitem\@bibitem
\let\st@lbibitem\@lbibitem
\ifnum\draftcontrol=1
  \def\@bibitem#1{%
    \st@bibitem{#1}\a@@label{#1}\ignorespaces}
  \def\@lbibitem[#1]#2{%
    \st@lbibitem[#1]{#2}\a@@label{#2}\ignorespaces}
  \def\a@@label#1{%
    \gdef\a@lab{\smash{\normalfont\small#1}}
    \ifvmode
      \if@inlabel
        \global\setbox\@labels\hbox{%
          \llap{\a@lab\let\a@lab\relax
                \kern\@totalleftmargin\kern\marginparsep}%
          \box\@labels}%
      \fi
    \fi}
\fi

\documentclass[12pt,letterpaper]{article}

\usepackage{amsmath,amssymb,array,calc,rotating,epsfig,psfrag}
\usepackage[nosort]{cite}

\ifnum\draftcontrol=1
\fi

\renewcommand\baselinestretch{1.25}
\renewcommand\section{\@startsection {section}{1}{\z@}%
                                   {-3.5ex \@plus -1ex \@minus -.2ex}%
                                   {2.3ex \@plus.2ex}%
                                   {\normalfont\large\bfseries}}
\renewcommand\subsection{\@startsection{subsection}{2}{\z@}%
                                   {-3.25ex\@plus -1ex \@minus -.2ex}%
                                   {1.5ex \@plus .2ex}%
                                   {\normalfont\normalsize\bfseries}}
\renewcommand\subsubsection{\@startsection{subsubsection}{3}{\z@}%
                                   {-3.25ex\@plus -1ex \@minus -.2ex}%
                                   {1.5ex \@plus .2ex}%
                                   {\normalfont\normalsize\it}}
\renewcommand\paragraph{\@startsection{paragraph}{4}{\z@}%
                                   {-3.25ex\@plus -1ex \@minus -.2ex}%
                                   {1.5ex \@plus .2ex}%
                                   {\normalfont\normalsize\bf}}

\numberwithin{equation}{section}

\def\ie{{\it i.e.}}

\def\revise#1       {\raisebox{-0em}{\rule{3pt}{1em}}%
                     \marginpar{\raisebox{.5em}{\vrule width3pt\
                     \vrule width0pt height 0pt depth0.5em
                     \hbox to 0cm{\hspace{0cm}{%
                     \parbox[t]{4em}{\raggedright\footnotesize{#1}}}\hss}}}}

\def\calf         {{\cal F}}

\def\calh         {{\cal H}}

\def\caln         {{\cal N}}
\def\calo         {{\cal O}}

\def\zet          {{\mathbb Z}}

\def\sqr#1#2{{\vcenter{\vbox{\hrule height.#2pt
 \hbox{\vrule width.#2pt height#1pt \kern#1pt
 \vrule width.#2pt}\hrule height.#2pt}}}}

\def\a{\alpha}
\def\ep{\epsilon}
\def\hep{\hat{\epsilon}}
\def\w{\omega}
\def\hh{\hat{h}}
\def\hw{\hat{\omega}}
\def\hb{\hat{b}}
\def\ha{\hat{a}}
\def\tH{\tilde{H}}
\def\H{\calh}

\catcode`\@=12

\begin{document}


\title{Inflation on the resolved warped deformed conifold}

\pubnum{%
UWO-TH-06/01
}
\date{January 2006}

\author{
Alex Buchel\\[0.4cm]
\it Department of Applied Mathematics\\
\it University of Western Ontario\\
\it London, Ontario N6A 5B7, Canada\\[0.2cm]
\it Perimeter Institute for Theoretical Physics\\
\it Waterloo, Ontario N2J 2W9, Canada\\
}

\Abstract{
Braneworld inflation on the resolved warped deformed conifold is
represented by the dynamics of a $D3$-brane probe with the world
volume of a brane spanning the large dimensions of the observable
Universe. This model was recently proposed as a string theory
candidate for slow-roll inflationary cosmology in hep-th/0511254.
During inflation, the scalar curvature of the Universe is determined
by the Hubble scale.  We argue that taking into account the curvature
of the inflationary Universe renders dynamics of the $D3$-brane
fast-roll deep inside the warped throat.
}


\makepapertitle

\body

\version\versionno

\section{Introduction}
Inflation \cite{in1,in2,in3} is an  attractive scenario which solves many important problems in cosmology. 
The basic idea of its simplest realization is that our Universe went through the stage of the
accelerated expansion driven by the potential energy of the slowly rolling inflaton field. 
In agreement with current observational data such a model naturally predicts a flat Universe 
and a scale invariant spectrum of density perturbations, provided the inflaton potential is sufficiently flat. 

The main problem of implementing inflation in string theory is to identify a string field 
 which has such a flat potential. In the original brane-world model scenario \cite{ddb1,ddb2,ddb3,ddb4}
( and its  more recent warped throat realization \cite{k2} ) the inflaton field is identified with
the geometric position of a $D3$-brane on the compactification manifold. The four-dimensional world
volume of a $D3$-brane is assumed to span the space-time directions of the observable Universe.      
In many supersymmetric string theory compactifications (a relevant example here is a 
warped deformed conifold \cite{ks}) a scalar field parameterizing position of a $D3$-brane on the transverse 
manifold is an exact modulus. A nontrivial potential for such a scalar can be generated by turning on 3-form 
fluxes. In the case of the warped deformed conifold one can turn on supersymmetry preserving fluxes \cite{grana}
(describing the baryonic branch deformation of the deformed conifold \cite{ks})
 to completely lift\footnote{Lifting a 3-brane flat directions with fluxes is 
a well-known phenomena. Indeed, a 3-brane on $AdS_5\times S^5$ 
has a six dimensional moduli space  which can be partially lifted \cite{bpp,cj} by turning on Pilch-Warner \cite{pw} 
fluxes.} a 3-brane flat directions \cite{dks}.  The resulting $D3$-brane potential on the resolved warped 
deformed conifold is rather flat. If $\Phi$ is a canonically normalized inflaton, \ie, the radial position 
of the $D3$-brane deep inside the warped throat of the resolved-deformed conifold, Dymarsky, Klebanov and Seiberg
(DKS) found \cite{dks} that $V(\Phi)$ satisfies 
\begin{equation}
\frac{d^2 V}{d\Phi^2}\sim U^4\  \Phi^{-6}\ln\Phi\,,\qquad \Phi^2\gg |U|\,,
\eqlabel{dksres}
\end{equation}           
where $U$ is a baryonic branch deformation parameter \cite{grana}. 
Potential \eqref{dksres} appear to be very promising for the slow roll inflation. 
Specifically, if a Hubble scale during inflation is $\calh$, using \eqref{dksres} the slow roll parameter $\eta$
is 
\begin{equation}
\eta\equiv \frac {1}{3\H^2}\frac{d^2 V}{d\Phi^2}\sim \H^{-2}\   U^4\  \Phi^{-6}\ln\Phi\,, 
\eqlabel{etadks}
\end{equation} 
which can be made arbitrarily small by considering inflation sufficiently far inside the warped throat.

Unfortunately, above arguments miss a crucial contribution to the probe brane potential which renders 
DKS inflationary model\footnote{We discuss here DKS inflation only deep inside 
warped throat. It might be possible to achieve slow-roll inflation 
at the bottom of the warped resolved deformed conifold of \cite{grana}.} 
unrealistic. Specifically, the main problem is that the potential 
\eqref{dksres} was evaluated assuming that the world-volume of the $D3$-brane is Minkowski.
As we already mentioned the world volume of the inflationary brane extends over the observable Universe.
During slow-roll inflation the Universe has a scalar curvature 
\begin{equation}
R_4=12 \H^2\,,
\eqlabel{r4}
\end{equation}  
and thus, the background geometry for the brane inflation is {\it not} a direct warped
product of a four-dimensional Minkowski space with the resolved-deformed conifold. Rather, it should be
a direct warped product of a four-dimensional de-Sitter space of curvature \eqref{r4}
with  the resolved-deformed conifold. Such clarification is very important because the scalar field 
representing $D3$-brane radial position on the conifold is a conformally coupled scalar \cite{k2,br}.
Thus, just from the conformal coupling to the curved background the DKS potential would receive an 
extra contribution (schematically)
\begin{equation}
V_{DKS}\Longrightarrow  V_{DKS}+\frac{1}{12}\ R_4 \Phi^2= \H^2\ \Phi^2\,,
\eqlabel{potdef}
\end{equation}   
which would lead to  
\begin{equation}
\eta_{DKS}\Longrightarrow  \eta_{DKS}+\frac 23\,.
\eqlabel{etadef}
\end{equation}   
Given that deep inside the warped throat $|\eta_{DKS}|\ll 1$, the {\it correct} slow roll parameter of the 
DKS inflation is actually $\frac 23$. 

In the rest of this paper we provide precise evaluation of the cosmological parameter $\eta$ in 
DKS inflationary model. First, building on \cite{bt,bds}, we derive supergravity equations of motion 
describing de-Sitter deformation of the resolved warped deformed conifold of \cite{grana}. 
In section 3 we find asymptotic solution to these background equations of motion. 
In section 4 we discuss $D3$ probe brane dynamics in de-Sitter deformed resolved warped 
deformed conifold. We conclude in section 5.    

\section{De-Sitter deformed resolved warped deformed conifold}
In this section we construct de-Sitter deformation of the supersymmetric resolved warped deformed 
conifold \cite{grana}. 

Following \cite{bt,bds}, we consider the following ansatz for the background (Einstein frame) metric
\begin{equation}
ds_{10}^2=c_1^2 \bigg(-dt^2+e^{2\H t}(d\bar{x})^2\bigg)+c_2^2 (dr)^2+c_3^2\sum_{i=1}^2 e_i^2
+c_4^2\sum_{i=1}^2\hep_i^2+c_5^2\hep_3^2\,, 
\eqlabel{10d}
\end{equation}
where the $S^2$ one-forms $e_i$ and the $S^3$ left-invariant one-forms $\ep_i$ are \cite{pt,grana,dks}
\begin{equation}
\begin{split}
&e_1=d\theta_1\,,\qquad e_2=-\sin\theta_1\ d\phi_1\,,\\
&\ep_1=\sin\psi\ \sin\theta_2\ d\phi_2+\cos\psi\ d\theta_2\,,\qquad
\ep_2=\cos\psi\ \sin\theta_2\  d\phi_2-\sin\psi\  d\theta_2\,,\\
& \ep_3=d\psi+\cos\theta_2\ d\phi_2\,,\\
&\hep_1=\ep_1-a\ e_1\,,\qquad \hep_2=\ep_2-a\ e_2\,,\qquad \hep_3=\ep_3+\cos\theta_1\ d\phi_1\,,
\end{split}
\eqlabel{1forms}
\end{equation}
with $c_i=c_i(r)$ and $a=a(r)$. The background fluxes take form \cite{pt}
\begin{equation}
\begin{split}
&H_3=d B_2\,,\\
&B_2=h_1\left(\ep_1\wedge \ep_2+e_1\wedge e_2\right)+h_2\left(\ep_1\wedge e_2-\ep_2\wedge e_1\right)
+\chi\left(-\ep_1\wedge \ep_2+e_1\wedge e_2\right)\,,\\
&F_3=P\ \hep_3\wedge \left(\ep_1\wedge \ep_2+e_1\wedge e_2\right)
+P\ d\biggl[b\ (\ep_1\wedge e_1+\ep_2\wedge e_2)\biggr]\,,\\
&F_5=\calf_5+\star_{10}\calf_5,\qquad \calf_5=K\ e_1\wedge e_2\wedge \ep_1\wedge \ep_2\wedge \ep_3\,,
\end{split}
\eqlabel{3form}
\end{equation}
where $h_i=h_i(r)$, $\chi=\chi(r)$, $b=b(r)$, $K=K(r)$, 
and $P=-\frac 14\a' M$ is determined by the number $M$ of 
fractional 3-branes. Finally, there is a dilaton $\phi\equiv \ln g_s=\phi(r)$, and we assume the
asymptotic string coupling to be one, \ie, $\phi\to 0$ as $r\to\infty$.

Type IIB supergravity equations of motion \cite{sugra} consist of 3-form Maxwell equations, 
dilaton equation, five-form Bianchi identity and the Einstein equations. 
The  five-form Bianchi identity can be integrated to yield
\begin{equation}
K=2P \left(h_1+b h_2\right)\,.
\eqlabel{5form}
\end{equation}  
The 3-form Maxwell equations reduce to the following coupled ODE's
\begin{equation}
\begin{split}
&0=b''+b'\ \left[\ln\frac{c_1^4c_5g_s}{c_2}\right]'
-\frac{b c_2^2 (c_4^2 g_s (c_3^2+2 c_4^2 a^2)+2 h_2^2)}{c_4^2 c_3^2 c_5^2 g_s}-2 \frac{c_2^2 h_2 h_1}{c_4^2 c_3^2 c_5^2 g_s}
\\
&+\frac{c_2^2 a (c_3^2+c_4^2 (a^2+1))}{c_3^2 c_5^2}\,,
\end{split}
\eqlabel{max1}
\end{equation}
\begin{equation}
\begin{split}
&0=h_1''+h_1'\left[\ln\frac{c_1^4c_5}{c_2 g_s}\right]'
+h_2'\biggl(a\left[\ln\frac{ac_4^2}{c_3^2}\right]'+\frac{c_4^2(1-a^2)}{c_3^2}\ 
a'\biggr)+\chi'\biggl(\left[\ln \frac{c_4^2}{c_3^2}\right]'
-2\frac{ac_4^2}{c_3^2}\ a'\biggr)\\
&+2 \frac{b^2 g_s c_2^2 a P^2 h_2 (c_3^2+c_4^2 (a^2+1))}{c_3^4 c_5^2 c_4^2}-\frac{b g_s c_2^2 P^2}{c_3^4 c_5^2 c_4^4} 
\biggl(-2 c_4^2 a h_1 (c_3^2+c_4^2 (a^2+1))\\
&+h_2 (c_4^4 (a^2+1)^2+c_3^2 (c_3^2+2 c_4^2 a^2))\biggr)
-\frac{h_1 g_s c_2^2 P^2 (c_4^4 (a^2+1)^2+c_3^2 (c_3^2+2 c_4^2 a^2))}{c_3^4 c_5^2 c_4^4}\\
&+\frac{h_2 c_2^2 a (c_3^2+c_4^2 (a^2+1))}{c_3^2 c_5^2}\,,
\end{split}
\eqlabel{max2}
\end{equation}
\begin{equation}
\begin{split}
&0=h_2''+h_2'\left[\ln\frac{c_1^4c_5}{c_2 g_s}\right]'
+h_1'\biggl(a\left[\ln\frac{ac_4^2}{c_3^2}\right]'+
\frac{c_4^2(1-a^2)}{c_3^2}\ 
a'\biggr)+\chi'\biggl(a\left[\ln \frac{c_3^2}{ac_4^2}\right]'\\
&+\frac{c_4^2(1+a^2)}{c_3^2}\ a'\biggr)
-2 \frac{b^2 g_s c_2^2 P^2 h_2 (c_3^2+2 c_4^2 a^2)}{c_3^4 c_5^2 c_4^2}-2 \frac{b g_s c_2^2 P^2}{c_3^4 c_5^2 c_4^2} 
\biggl(-a h_2 (c_3^2+c_4^2 (a^2+1))\\
&+h_1 (c_3^2+2 c_4^2 a^2)\biggr)+2 \frac{h_1 g_s c_2^2 P^2 a (c_3^2+c_4^2 (a^2+1))}{c_3^4 c_5^2 c_4^2}
-\frac{h_2 c_2^2 (c_3^2+2 c_4^2 a^2)}{c_3^2 c_5^2}\,,
\end{split}
\eqlabel{max3}
\end{equation}
\begin{equation}
\begin{split}
&0=\chi''+\chi'\left[\ln\frac{c_1^4c_5}{c_2 g_s}\right]'
+h_1'\biggl(\left[\ln\frac{c_4^2}{c_3^2}\right]'-2
\frac{a c_4^2}{c_3^2}\ 
a'\biggr)+h_2'\biggl(a\left[\ln\frac{ac_4^2}{c_3^2}\right]'
-\frac{c_4^2(1+a^2)}{c_3^2}\ a'\biggr)\\
&+2 \frac{b^2 g_s c_2^2 a P^2 h_2 (c_3^2-c_4^2 (1-a^2))}{c_3^4 c_5^2 c_4^2}-
\frac{b g_s c_2^2 P^2}{c_3^4 c_5^2 c_4^4} \biggl(h_2 ((c_3^2+c_4^2 a^2)^2-c_4^4)\\
&-2 c_4^2 a h_1 (c_3^2-c_4^2+c_4^2 a^2)\biggr)
-\frac{h_1 g_s c_2^2 P^2 ((c_4^2 a^2+c_3^2)^2-c_4^4)}{c_3^4 c_5^2 c_4^4}+\frac{h_2 c_2^2 a (c_3^2-c_4^2 (1-a^2))}{c_3^2 c_5^2}\,,
\end{split}
\eqlabel{max4}
\end{equation}
\begin{equation}
\begin{split}
&0=-\chi' (c_4^4 (1-a^2)^2+c_3^2 (c_3^2+2 c_4^2 a^2))
+h_1' ((c_3^2+c_4^2 a^2)^2-c_4^4)+ 2 c_4^2 a h_2' (c_3^2-c_4^2+c_4^2 a^2)\,.
\end{split}
\eqlabel{max5}
\end{equation}
It is straightforward to verify that the first order constraint \eqref{max5} is consistent with \eqref{max1}-\eqref{max4}.
The dilaton equation is  
\begin{equation}
\begin{split}
&0=g_s''-\frac{(g_s')^2}{g_s}+g_s' \left[\ln\frac{c_1^4c_3^2c_4^2c_5}{c_2}\right]'
-\frac{g_s^2 P^2 (b')^2}{c_3^2 c_4^2}+\frac{ (\chi')^2}{2c_3^4 c_4^4} 
\biggl((c_4^2 a^2+c_3^2)^2-c_4^4 (-1+2 a^2)\biggr)\\
&-\frac{\chi'}{c_3^4 c_4^4}\biggl(c_3^2-c_4^2(1- a^2)\biggr) \biggl(2 c_4^2 a h_2'
+h_1' (c_3^2+c_4^2(1+ a^2))\biggr) -\frac{g_s^2 c_2^2 P^2}{2c_3^4 c_5^2 c_4^4}  \biggl(
\\
&2 c_4^2 (c_3^2+2 c_4^2 a^2) (b-a)^2
+c_3^4+c_4^4 (1-a^2) (3 a^2+1)+4 b a c_4^4 (a^2-1)\biggr)+\frac{c_2^2 h_2^2}{c_3^2 c_4^2 c_5^2}\\
&+\frac{(c_3^2+2 c_4^2 a^2) (h_2')^2}{c_3^4 c_4^2}
+\frac{(h_1')^2}{2c_3^4 c_4^4} \biggl(c_4^4 (a^2+1)^2+c_3^2 (c_3^2+2 c_4^2 a^2)\biggr) \\
&+2 \frac{ah_1'  h_2'}{c_3^4 c_4^2} 
\biggl(c_3^2+c_4^2 (a^2+1)\biggr)\,.
\end{split}
\eqlabel{dil}
\end{equation}
The Einstein equations are
\begin{equation}
\begin{split}
&0=c_1''+\frac{3(c_1')^2}{c_1}+c_1' \left[\ln\frac{c_3^2 c_4^2 c_5}{c_2}\right]' -\frac{c_1(\chi')^2}{8c_4^4 c_3^4 g_s}  \biggl((c_4^2 a^2+c_3^2)^2+c_4^4 (1-2 a^2)\biggr)\\
&+\frac{c_1\chi'}{4c_4^4 c_3^4 g_s}  \biggl(c_3^2+c_4^2 (a^2-1)\biggr) \biggl(2 c_4^2 a h_2'+h_1' (c_3^2+c_4^2 (a^2+1))\biggr)   
-\frac{c_1 P^2 g_s(b')^2}{4c_4^2 c_3^2}\\
&-\frac{c_1(h_1')^2}{8c_4^4 c_3^4 g_s}  \biggl(c_4^4 (a^2+1)^2+c_3^2 (c_3^2+2 c_4^2 a^2)\biggr) 
 -\frac{c_1(h_2')^2}{4c_4^2 g_s c_3^4}  \biggl(c_3^2+2 c_4^2 a^2\biggr) \\
& -\frac{h_1'h_2'c_1a}{2c_4^2 g_s c_3^4}   \biggl(c_3^2+c_4^2 (a^2+1)\biggr)
 -\frac{g_sc_2^2 c_1 P^2}{8c_4^4 c_3^4 c_5^2}  \biggl(2 c_4^2 (c_3^2+2 c_4^2 a^2) (b-a)^2\\
&+ c_3^4+c_4^4 (1-a^2) (3 a^2+1)+4 b a c_4^4 (a^2-1)\biggr)-\frac{c_2^2 c_1 h_2^2}{4c_4^2 c_3^2 c_5^2g_s} 
-\frac{c_2^2 c_1 P^2 (h_1+b h_2)^2}{c_3^4 c_4^4 c_5^2}\\
&-3 \frac{c_2^2 \H^2}{c_1}\,,
\end{split}
\eqlabel{c1}
\end{equation}
\begin{equation}
\begin{split}
&0=c_3''+\frac{(c_3')^2}{c_3} +c_3' \bigg[\ln\frac{c_1^4 c_4^2 c_5}{c_2}\biggr]'+ \frac{(\chi')^2}{8c_4^4 g_s c_3^3} \biggl(3 c_4^4 (1-a^2)^2+c_3^2 (2 c_4^2 a^2
-c_3^2)\biggr) \\
&+\frac{\chi'}{4c_4^4 g_s c_3^3} \biggl(h_1' (c_3^2 (-2 c_4^2 a^2+c_3^2)-3 c_4^4 (a^4-1))-2 c_4^2 a h_2' (3 c_4^2 (a^2-1)+c_3^2)\biggr) 
\\
&+ \frac{(h_1')^2}{8c_4^4 g_s c_3^3} \biggl(3 c_4^4 (a^2+1)^2+c_3^2 (2 c_4^2 a^2-c_3^2)\biggr)
+\frac{(h_2')^2}{4c_4^2 g_s c_3^3} (c_3^2+6 c_4^2 a^2)\\
&+ \frac {h_1'h_2'a}{2c_4^2 g_s c_3^3}  \biggl(c_3^2+3 c_4^2 (1+a^2)\biggr) 
+ \frac{g_s P^2(b')^2}{4c_4^2 c_3}+\frac{(a')^2c_4^2}{2 c_3} -\frac{g_sc_2^2 P^2}{8c_4^4 c_3^3 c_5^2}  \biggl(
\\
&-2 c_4^2 (c_3^2+6 c_4^2 a^2) (b-a)^2
-12 b a c_4^4 (a^2-1)+3 c_4^4 (a^2-1) (3 a^2+1)+c_3^4\biggr)\\
&+ \frac{c_2^2 h_2^2}{4c_4^2 c_3 c_5^2g_s}+\frac{P^2  c_2^2 (h_1+b h_2)^2}{c_4^4 c_3^3 c_5^2}
+\frac{c_5^2  c_2^2}{2c_4^2 c_3^3} \biggl(c_4^2 (a^2-1)^2+a^2 c_3^2\biggr)+ \frac{c_4^2 c_2^2 a^2}{2c_3 c_5^2}\\
& -\frac{c_2^2 (a^2+1)}{c_3}\,,
\end{split}
\eqlabel{c3}
\end{equation}
\begin{equation}
\begin{split}
&0=c_4''+\frac{(c_4')^2}{c_4}+c_4' \left[\ln\frac{c_1^4 c_3^2 c_5}{c_2}\right]'
+ \frac{(\chi')^2}{8c_3^4 g_s c_4^3} \biggl(c_3^2 (2 c_4^2 a^2+3 c_3^2)-c_4^4 (a^2-1)^2\biggr)\\
&-\frac{\chi'}{4c_3^4 g_s c_4^3} \biggl(h_1' (c_3^2 (2 c_4^2 a^2+3 c_3^2)+c_4^4 (1-a^4))+2 c_4^2 h_2' a (c_3^2+c_4^2 (1-a^2))\biggr)
\\
& -\frac{(h_1')^2}{8c_3^4 g_s c_4^3} \biggl(c_4^4 (a^2+1)^2-c_3^2 (2 c_4^2 a^2+3 c_3^2)\biggr)
-\frac{(h_2')^2}{4c_4 c_3^4 g_s} \biggl(2 c_4^2 a^2-c_3^2\biggr) \\
&+\frac{h_1'h_2'a}{2c_4 c_3^4 g_s}  \biggl(c_3^2-c_4^2 (1+a^2)\biggr)+ \frac{P^2 g_s(b')^2}{4c_4 c_3^2}
-\frac{c_4^3(a')^2}{2c_3^2}+ \frac{g_sc_2^2 P^2}{8c_3^4 c_4^3 c_5^2}  \biggl(\\
&2 c_4^2 (-2 c_4^2 a^2+c_3^2) (b-a)^2-4 b a c_4^4 (a^2-1)+c_4^4 (a^2-1) (3 a^2+1)+3 c_3^4\biggr)\\
&+\frac{c_2^2 h_2^2}{4c_4 c_3^2 c_5^2g_s}+\frac{P^2  c_2^2 (h_1+b h_2)^2}{c_3^4 c_4^3 c_5^2}
+\frac{c_5^2 c_2^2}{2c_3^2 c_4^3} \biggl(c_4^2 a^2+c_3^2\biggr)-\frac{c_2^2}{c_4} - \frac{c_4^3 c_2^2 a^2}{2c_3^2 c_5^2}\,,
\end{split}
\eqlabel{c4}
\end{equation}
\begin{equation}
\begin{split}
&0=c_5''+c_5' \left[\ln\frac{c_1^4 c_3^2 c_4^2}{c_2}\right]' 
-\frac{c_5(\chi')^2}{8g_s c_4^4 c_3^4}  \biggl(c_4^4 (a^2-1)^2+c_3^2 (c_3^2+2 c_4^2 a^2)\biggr)
\\
&+ \frac{c_5\chi'}{4g_s c_4^4 c_3^4}  \biggl(c_3^2+c_4^2 (a^2-1)\biggr) 
\biggl(2 c_4^2 a h_2'+h_1' (c_3^2+c_4^2 (a^2+1))\biggr) -\frac{c_5(h_1')^2}{8g_s c_4^4 c_3^4}  \biggl(\\
&c_4^4 (a^2+1)^2+c_3^2 (c_3^2+2 c_4^2 a^2)\biggr) 
-\frac{c_5(h_2')^2}{4c_4^2 g_s c_3^4} \biggl(c_3^2+2 c_4^2 a^2\biggr)  - \frac{c_5 ah_1'h_2'}{2c_4^2 g_s c_3^4}  \biggl(c_3^2+c_4^2 (1+a^2)\biggr)\\
&  - \frac{g_s c_5 P^2(b')^2}{4c_4^2 c_3^2} +\frac{3g_sc_2^2 P^2}  {8c_5 c_3^4 c_4^4}  \biggl(
2 c_4^2 (c_3^2+2 c_4^2 a^2) (b-a)^2+4 b a c_4^4 (a^2-1)\\
&-c_4^4 (a^2-1) (3 a^2+1)+c_3^4\biggr)+ \frac{3 c_2^2 h_2^2}{4c_4^2 c_3^2 c_5g_s}+\frac{P^2  c_2^2 (h_1+b h_2)^2}{c_5 c_3^4 c_4^4}
\\
&+ \frac{c_2^2 c_4^2 a^2}{c_3^2c_5} -\frac{ c_2^2c_5^3}{2c_4^4 c_3^4} \biggl(c_4^4 (a^2-1)^2+c_3^2 (c_3^2+2 c_4^2 a^2)\biggr)\,,
\end{split}
\eqlabel{c5}
\end{equation}
\begin{equation}
\begin{split}
&0=a''+a' \left[\ln\frac{c_1^4 c_4^4 c_5}{c_2}\right]' -\frac{a(\chi')^2 }{g_s c_3^2 c_4^4}\biggr(c_3^2+c_4^2 (a^2-1)\biggr)
+ \frac{\chi'}{g_s c_3^2 c_4^4}\biggl(h_2' (c_4^2 (3 a^2-1)+c_3^2)\\
&+2 h_1' a (c_4^2 a^2+c_3^2)\biggr) -\frac{a(h_1')^2}{g_s c_3^2 c_4^4} \biggl(c_3^2+c_4^2 (1+a^2)\biggr)  -2 \frac{a(h_2')^2}{g_s c_3^2 c_4^2}\\
&-\frac{h_1'h_2'}{g_s c_3^2 c_4^4} \biggl(c_3^2+c_4^2 (3 a^2+1)\biggr) -\frac{P^2 c_2^2 g_s (a-b)}{c_3^2 c_4^4 c_5^2} \biggl(c_4^2 (a^2-2 b a+1)+c_3^2\biggr) 
\\
&-\frac{c_2^2 ac_5^2}{c_3^2 c_4^4} \biggl(c_3^2+c_4^2 (a^2-1)\biggr)+\frac{c_2^2 a(2 c_5^2-c_4^2)}{c_4^2 c_5^2}\,, 
\end{split}
\eqlabel{a}
\end{equation}
\begin{equation}
\begin{split}
&0=4 c_3^2 g_s^2 c_5 c_4^2 \biggl(c_5 c_1^2 c_4^2 (c_3')^2+c_5 c_3^2 c_1^2 (c_4')^2+4 c_5 c_3 c_1^2 c_4 c_3' c_4'
+8 c_5 c_3^2 c_1 c_1'c_4' c_4+8 c_5 c_3 c_1 c_4^2 c_1' c_3'\\
&+6 c_5 c_3^2 c_4^2 (c_1')^2+4 c_4^2 c_3^2 c_1 c_1' c_5'
+2 c_4^2 c_3 c_1^2 c_5' c_3'+2 c_3^2 c_1^2 c_5' c_4' c_4\biggr)-c_1^2 c_5^2 g_s \biggl(c_4^4 (a^2-1)^2\\
&+c_3^2 (c_3^2+2 c_4^2 a^2)\biggr) (\chi')^2 
+2 c_1^2 c_5^2 g_s \biggl(c_3^2-c_4^2 (1-a^2)\biggr) \biggl(2 c_4^2 a h_2'+h_1' (c_3^2+c_4^2 (a^2+1))\biggr) \chi' \\
&-c_1^2 c_5^2 g_s \biggl(c_4^4 (a^2+1)^2+c_3^2 (c_3^2+2 c_4^2 a^2)\biggr) (h_1')^2 
-2 c_1^2 c_4^2 c_5^2 g_s \biggl(c_3^2+2 c_4^2 a^2\biggr) (h_2')^2\\
&-4 c_1^2 c_4^2 c_5^2 h_1'h_2' g_s a \biggl(c_3^2+c_4^2 (1+a^2)\biggr)
-c_3^4 c_4^4 c_5^2 c_1^2 (g_s')^2 -2 g_s^3 c_1^2 P^2 c_4^2 c_3^2 c_5^2 (b')^2\\
&-2 g_s^2 c_3^2 c_4^6 c_5^2 c_1^2 (a')^2  -4 g_s^2 c_1^2 c_2^2 c_4^2 c_3^2 \biggl(c_3^2+c_4^2 (1+ a^2)\biggr) c_5^2 
+g_s^2 c_1^2 c_2^2 \biggl(c_4^4 (a^2-1)^2\\
&+c_3^2 (c_3^2+2 c_4^2 a^2)\biggr) c_5^4+ 2 g_s c_1^2 c_2^2 c_4^2 c_3^2 \biggl(h_2^2+g_s c_4^4 a^2\biggr) 
+4 P^2 c_2^2 c_1^2 (h_1+b h_2)^2 g_s^2\\
&+ P^2 g_s^3 c_1^2 c_2^2 \biggl(
2 c_4^2 (c_3^2+2 c_4^2 a^2) (b-a)^2+4 b a c_4^4 (a^2-1)-c_4^4 (a^2-1) (3 a^2+1)+c_3^4\biggr)\\
&-24 g_s^2 c_2^2 c_3^4 c_4^4 c_5^2 \H^2\,.
\end{split}
\eqlabel{cc}
\end{equation}
It is straightforward to verify that the constraint \eqref{cc} is consistent with \eqref{max1}-\eqref{max4}, \eqref{max5},
 \eqref{dil}, \eqref{c1}-\eqref{a}. Notice that the Hubble scale $\H$ enters only in \eqref{c1} and in the constraint equation \eqref{cc}.

As a consistency check, we verified that the supersymmetric baryonic branch deformation of the warped deformed conifold of  \cite{grana}
is indeed a solution of derived supergravity equations of motion with $\calh=0$. Comparison with computation of  \cite{dks}
is achieved by parameterizing 
\begin{equation}
\begin{split}
&c_1=e^{-\phi/4}H^{-1/4}\,,\qquad c_2=c_5=e^{-\phi/4} v^{-1/2}e^{x/2}\,,\\
&c_3=^{-\phi/4} e^{g/2}e^{x/2}\,,\qquad  c_4=^{-\phi/4} e^{-g/2}e^{x/2}\,.
\end{split}
\eqlabel{dkstrans}
\end{equation}
where $\{\phi,H,v,g,x\}$ are functions of a radial variable, see Eq.~(12.2) of \cite{dks}. 
We further reproduced (again setting $\H=0$) the warped deformed conifold solution of Klebanov and Strassler (KS)\cite{ks}.
We verified the map between KS background parametrization and the parametrization of the $\zet_2$-symmetric solution of \cite{grana}, 
presented in \cite{dks}.

\section{Asymptotic solution}
In this section we discuss asymptotic infrared (IR) and ultraviolet (UV) solution to 
the de-Sitter deformed resolved warped deformed conifold
supergravity equations of motion derived in section 2. We find that the IR (the bottom of the warped deformed 
throat) asymptotic is a smooth deformation\footnote{In other 
words, there is no obstruction in the IR for the introduction of
 the Hubble parameter.} of the IR solution of \cite{dks} by turning on a nonzero 
Hubble scale $\calh$. Thus while  in \cite{dks} the IR geometry is that of $R^{3,1}\times R^3\times S^3$,
in our case it is  $dS_4\times R^3\times S^3$. The UV asymptotic describes a highly warped 
region of the geometry where following the proposal of \cite{dks} we study $D3$-brane inflation.

\subsection{IR solution}
We assume the same IR boundary conditions as for the warped deformed conifold \cite{ks}, \ie, we assume that the $S^2$ of the conifold shrinks to zero size 
in a smooth way, while all the other warp factors (including  the $S^3$ radius of the conifold) remain finite. It is easy to verify that 
such  boundary conditions guarantee singularity-free geodesically complete space-time. Moreover, as in \cite{ks}, the curvature invariants of the background can 
then be made small in string units by choosing fluxes in the IR to be large. So, we search for a small-$r$ solution to  \eqref{max1}-\eqref{max4}, \eqref{max5},
 \eqref{dil}, \eqref{c1}-\eqref{a} subject to the following $r\to 0$ boundary condition on the geometry
\begin{equation}
ds_{10}^2\Rightarrow H_0^{-1/2}\biggl(-dt^2+e^{2\H t}(d\bar{x})^2\biggr)+H_0^{1/2}
\biggl((dr)^2+r^2\sum_{i=1}^2 e_i^2+\w_0^2(g_5^2+2g_3^2+2g_4^2)\biggr)\,,
\eqlabel{irboundary}  
\end{equation}
where the one-forms $\{g_3,g_4,g_5\}$
\begin{equation}
g_5\equiv \hep_3\,,\qquad g_3=\frac{e_2+\ep_2}{\sqrt{2}}\,,\qquad g_4=\frac{e_1+\ep_1}{\sqrt{2}}\,,
\eqlabel{gdef}
\end{equation}
are defined so as to agree with \cite{ks}. The radius-squared of $S^3$ is $R^2=2H_0^{1/2}\w_0^2$, 
while the $S^2$ parameterized by $e_i$ 
smoothly shrinks to zero size.
Notice that choosing the IR asymptotic of 
$c_2$ to satisfy \eqref{irboundary} we fixed the rescaling symmetry of the radial coordinate $r$.
Introducing 
\begin{equation}
c_1\equiv \tH^{-1/4}\,,\qquad c_2=\tH^{1/4}\,,\qquad c_3=\tH^{1/4}w_3\,, \qquad c_4=\tH^{1/4}w_4\,, \qquad c_5=\tH^{1/4}w_5\,,
\eqlabel{irdef} 
\end{equation}
and setting\footnote{We can not simultaneously set the string coupling to one both in the 
UV and in the IR. However, doing so separately in the IR and the UV is possible. 
Extension to $g_s(r=0)\ne 1$ is trivial.} $g_s(r=0)=1$, we find that  solution  to \eqref{max1}-\eqref{max4}, \eqref{max5},
 \eqref{dil}, \eqref{c1}-\eqref{a} subject to \eqref{irboundary} takes the form
\begin{equation}
\begin{split}
&\tH=H_0+\sum_{i=1}^\infty\ H_{2i}\ r^{2i}\,,\qquad w_3= r\biggl(1+\sum_{i=1}^\infty w_{3,2i}\ r^{2i}\biggr)\,,\qquad 
w_4= \w_0+\sum_{i=1}^\infty w_{4,2i}\ r^{2i}\,,\\
&w_5= \w_0+\sum_{i=1}^\infty w_{5,2i}\ r^{2i}\,,\qquad a=-1+\sum_{i=1}^\infty a_{2i}\ r^{2i}\,,\qquad 
g_s=1+\sum_{i=1}^\infty g_{s,2i}\ r^{2i}\,,\\
&b= \sum_{i=0}^\infty b_{2i}\ r^{2i}\,,\qquad h_1=P\ r\ \sum_{i=0}^\infty h_{1,2i}\ r^{2i}\,,\qquad 
h_2=P\ r\ \sum_{i=0}^\infty h_{2,2i}\ r^{2i}\,,
\end{split}
\eqlabel{expa}
\end{equation}
with $\chi'$ determined algebraically from \eqref{max5}.
The most general solution is characterized by seven parameters\footnote{For $\H =0$ one can remove 
$H_0$ by rescaling the four-dimensional space-time coordinates \cite{grana}. 
With $\H \ne0$, such a rescaling is no longer possible. }: $\{H_0,\w_0,h_{1,0},h_{1,2},h_{2,2},a_2,b_2\}$.
We present some  of the terms in the perturbative solution \eqref{expa}
\begin{equation}
\begin{split}
&b_0=-1\,,\qquad h_{2,0}=h_{1,0}\,,\\
&H_2=-\frac{P^2}{12\w_0^6}-\frac{P^2 h_{1,0}^2}{4\w_0^4}-\frac{P^2 b_2^2}{\w_0^2}-3 P^2 (h_{1,2}-h_{2,2})^2-2 H_0^2 \H^2\,,\\
&g_{s,2}=\frac{P^2}{12H_0 \w_0^6}- \frac{P^2 h_{1,0}^2}{4H_0 \w_0^4}+\frac{P^2 b_2^2}{H_0 \w_0^2}-3\frac{ P^2 }{H_0}(h_{1,2}-h_{2,2})^2\,,\\
&w_{3,2}=\frac{P^2}{24 H_0 \w_0^6}+\frac{ P^2 h_{1,0}^2}{24H_0 \w_0^4}- \frac{1}{24H_0 \w_0^2}(4 P^2 b_2^2+H_0)
-3\frac{ P^2}{2H_0}(h_{1,2}-h_{2,2})^2-\frac 12 a_2^2 \w_0^2\,,\\
&w_{4,2}=-\frac{P^2}{48 H_0 \w_0^5}+\frac{7 P^2 h_{1,0}^2}{240H_0 \w_0^3}
- \frac{1}{120H_0 h_{1,0} \w_0}(h_{1,0} (14 P^2 b_2^2+5 H_0)+24 P^2 b_2 (h_{1,2}\\
&-h_{2,2}))+\frac{\w_0}{20H_0 h_{1,0}} (27 P^2 h_{1,0} (h_{1,2}-h_{2,2})^2+H_0 (8 a_2 h_{1,0}+10 h_{2,2}+13 H_0 h_{1,0} \H^2))\\
&+\frac{a_2\w_0^3}{10h_{1,0}} (a_2 h_{1,0}+12 h_{1,2}-12 h_{2,2}) -6 \frac{ a_2^2\w_0^5}{5h_{1,0}} (h_{1,2}-h_{2,2})\,, \\
&w_{5,2}=-\frac{P^2}{12H_0 \w_0^5}-11\frac{P^2 h_{1,0}^2}{60H_0 \w_0^3}+\frac{1}{15H_0 h_{1,0} \w_0} 
(h_{1,0}(5 H_0+11 P^2 b_2^2) +6 P^2 (h_{1,2}-h_{2,2}) b_2\\
&)+\frac{\w_0}{5H_0 h_{1,0}} (H_0 (-5 h_{2,2}-4 a_2 h_{1,0}+H_0 h_{1,0} \H^2)+9 P^2 h_{1,0} (h_{1,2}-h_{2,2})^2)
\\ 
&+4\frac{ a_2\w_0^3}{5h_{1,0}} (-3 h_{1,2}+3 h_{2,2}+a_2 h_{1,0}) +12 \frac{a_2^2\w_0^5}{5h_{1,0}} (h_{1,2}-h_{2,2}) \,.
\end{split}
\eqlabel{solir}
\end{equation}
As in \cite{grana}, we expect that decoupling of the asymptotic Minkowski region as $r\to\infty$ 
would constrain some of these IR parameters. Specifically, in supersymmetric case the presence of the 
AdS-like boundary determines the size of $S^3$ (corresponding to our $\w_0$) 
in terms of the baryonic branch parameter (corresponding to our $a_2$). In our case, the presence of the 
boundary (more detailed analysis are presented below) as $r\to \infty$ require 
\begin{equation}
\tH\to 0\,,\qquad  \tH^{1/4}w_i\to  \calo(\ln^{1/4}r)\,,\qquad a\to 0\,.
\eqlabel{bound}
\end{equation} 
Thus we expect at least\footnote{Some of constraints \eqref{bound} might be redundant. 
In particular, de-Sitter deformation of the KS cascading gauge theory baryonic branch 
discussed here might ``turn on''  extra relevant operators.  
It would be very interesting to analyze gauge/string theory correspondence in 
this setup along the lines of \cite{aby}.} two independent IR parameters. From the perspective of the dual gauge theory 
we have also at least two 
independent physical parameters: the ratio of the gauge theory strong coupling scale and  the Hubble scale, and the
baryonic branch deformation parameter.

\subsection{UV solution}
As in \cite{ks} and \cite{grana} we choose the radial gauge $c_2=c_5$. 
Asymptotically as $r\to \infty$ we expect to recover KS solution \cite{ks}. We find it convenient to introduce 
a new radial coordinate 
\begin{equation}
x=e^{-r/3}\,,
\eqlabel{def}
\end{equation}
which maps the latter boundary to $x\to 0$. 
Introduce\footnote{Again, $\chi'$ is determined algebraically from \eqref{max5}.} 
\begin{equation}
\begin{split}
&c_1=\tH^{-1/4}\,,\qquad c_3=\tH^{1/4}w_3\,,\qquad c_4=\tH^{1/4}w_4\,,\qquad c_2=c_5=\tH^{1/4}w_5\,,\\
&\tH=\frac{4P^2}{\lambda^4}\ x^4\ \hh\,,\qquad w_3=\lambda\ x^{-1}\ \hw_3\,,\qquad   w_4=\lambda\ x^{-1}\ \hw_4\,,\qquad
w_5=\sqrt{\frac 23}\lambda\ x^{-1}\ \hw_5\,,\\
&h_1=P \hh_1\,,\qquad h_2=P x^3\ \hh_2\,,\qquad b=x^3\ \hb\,,\qquad a=x^3\ \ha\,,
\end{split}
\eqlabel{uvdef}
\end{equation} 
where 
\begin{equation}
\lambda=2^{-4/3}\ep^{2/3}\,,
\eqlabel{ldef}
\end{equation}
is related to the conifold deformation parameter $\ep$ \cite{ks}.
To have leading KS asymptotics as  $x\to 0$  we require \cite{dks}
\begin{equation}
\begin{split}
&\hh\to \frac{3}{32}\left(-12\ln x-1\right)\,,\qquad \hw_3\to 1\,,\qquad \hw_4=\to 1\,,\qquad \hw_5\to 1\,,\\
&\hh_1\to -1-3\ln x\,,\qquad \hh_2\to -2-6\ln x\,,\qquad \hb\to 6\ln x\,,\qquad \ha\to -2\,,\qquad g_s\to 1\,. 
\end{split}
\eqlabel{lead}
\end{equation}
Notice that we set the asymptotic string coupling to one. 
We find the solution to 
\eqref{max1}-\eqref{max4}, \eqref{max5},
 \eqref{dil}, \eqref{c1}-\eqref{a} subject to \eqref{lead}
to take a form of a double series expansion
\begin{equation}
\begin{split}
&\hh=\sum_{i=0}^{\infty}\ x^{2i}\sum_{j=0}^{2i}\ \hh_{(i,j)}\ln^j x\,,
\qquad   \hw_3=\sum_{i=0}^{\infty}\ x^{2i}\sum_{j=0}^{2i}\ \hw_{3,(i,j)}\ln^j x\,,\\
\end{split}
\eqlabel{soluvan1}
\end{equation}
\begin{equation}
\begin{split}
&\hw_4=\sum_{i=0}^{\infty}\ x^{2i}\sum_{j=0}^{2i}\ \hw_{4,(i,j)}\ln^j x\,,
\qquad \hw_5=\sum_{i=0}^{\infty}\ x^{2i}\sum_{j=0}^{2i}\ \hw_{5,(i,j)}\ln^j x\,,\\
\end{split}
\eqlabel{soluvan2}
\end{equation}
\begin{equation}
\begin{split}
&\hh_1=\sum_{i=0}^{\infty}\ x^{2i}\sum_{j=0}^{2i}\ \hh_{1,(i,j)}\ln^j x\,,
\qquad \hh_2=\sum_{i=0}^{\infty}\ x^{2i}\sum_{j=0}^{2i}\ \hh_{2,(i,j)}\ln^j x\,,\\
\end{split}
\eqlabel{soluvan3}
\end{equation}
\begin{equation}
\begin{split}
&\hb=\sum_{i=0}^{\infty}\ x^{2i}\sum_{j=0}^{2i}\ \hb_{(i,j)}\ln^j x\,,
\qquad \ha=\sum_{i=0}^{\infty}\ x^{2i}\sum_{j=0}^{2i}\ \ha_{(i,j)}\ln^j x\,,\\
&g_s=\sum_{i=0}^{\infty}\ x^{2i}\sum_{j=0}^{2i}\ g_{s,(i,j)}\ln^j x\,.
\end{split}
\eqlabel{soluvan4}
\end{equation}
In order to match (in the $\H=0$ limit) the probe brane potential computed in \cite{dks} we need to find perturbative solution at least up to order 
$i=4$. To this order we find the solution to be characterized by seven parameters. 
Upon appropriately adjusting these parameters, we verified that in the $\H=0$ supersymmetric limit our expansion 
agrees precisely with the asymptotic expansion presented in \cite{grana}.  
As the coefficients of \eqref{soluvan1}-\eqref{soluvan4} are rather involved, we present   
here only the first nontrivial correction beyond the KS asymptotic \eqref{lead}. We find
\begin{equation}
\begin{split}
&\hh=-\frac{3}{32}-\frac 98 \ln x+\frac{27P^2 \H^2}{128\lambda^2}\ x^2\  (-96 \ln x+144 \ln^2 x+71)+\calo\left(x^4\ln^3 x\right)\,,\\
&\hw_3=1+x^2\ \biggl(\hw_{3,(2,1)} \ln x+\frac{117 P^2 \H^2}{16\lambda^2}-\frac 14 \ha_{(2,0)}\biggr)+\calo\left(x^4\ln^2 x\right)\,,\\
&\hw_4= 1+x^2\ \biggl(\ln x\biggl[\frac{27 P^2 \H^2}{2\lambda^2}-\hw_{3,(2,1)}\biggr]-\frac{207 P^2 \H^2}{16\lambda^2}+\frac 14 \ha_{(2,0)}\biggr)+\calo\left(x^4\ln^2 x\right)\,,\\
&\hw_5=1+\frac{9 P^2 \H^2}{16\lambda^2}\ x^2\  \biggl(12 \ln x+1\biggr) +\calo\left(x^4\ln^2 x\right)\,,\\
&\hh_1=-1-3 \ln x-\frac{27 P^2 \H^2}{8\lambda^2}\ x^2\  \biggl(12 \ln x-17\biggr)+\calo\left(x^4\ln^3 x\right)\,,\\
&\hh_2=-2-6 \ln x+\frac{27 P^2 \H^2}{32\lambda^2}\ x^2\  \biggl(144 \ln^2 x-72 \ln x+245\biggr)+\calo\left(x^4\ln^4 x\right)\,,\\
&\hb=6 \ln x-\frac{27 P^2 \H^2}{32\lambda^2}\ x^2\  \biggl(144 \ln^2 x-264 \ln x+157\biggr)+\calo\left(x^4\ln^4 x\right)\,,\\
&\ha=-2+x^2\ \biggl(\ln x\biggl[\frac{27 P^2 \H^2}{\lambda^2} -4 \hw_{3,(2,1)}\biggr]+\ha_{(2,0)}\biggr)+\calo\left(x^4\ln^3 x\right)\,,\\
&g_s=1-\frac{27 P^2 \H^2}{\lambda^2}\ x^2+\calo\left(x^4\ln^3 x\right)\,.
\end{split}
\eqlabel{order2}
\end{equation}
Notice that to order $i=1$ above only two UV parameters appear. The supersymmetric limit of \cite{grana} (to this order) is reproduced 
setting $\H=0$ and 
\begin{equation}
\hw_{3,(2,1)}=-\frac 34 \ha_{(2,0)}\,.
\eqlabel{w321}
\end{equation}

\section{$D3$-brane inflation on resolved warped deformed conifold}
Effective action of a $D3$ probe brane on the background \eqref{10d} moving along the radial direction takes form \cite{br,bi0}
\begin{equation}
S_{eff}=T_3\ \int\ d^4\xi\sqrt{-\gamma}\biggl(\ \frac 12 c_1^2 c_2^2\left(\frac{dr}{dt}\right)^2+C_4-c_1^4\biggr)\,,
\eqlabel{effact}
\end{equation} 
where $T_3$  is the 3-brane tension, $\gamma_{\mu\nu}$ is the metric of a four-dimensional de-Sitter space
\begin{equation}
\gamma_{\mu\nu}\ d\xi^\mu d\xi^\nu\equiv -dt^2+e^{2\H t}(d\bar{x})^2\,,
\end{equation}
and the four-form potential $C_4$ satisfies
\begin{equation}
\frac{dC_4}{dr}=K\ \frac{c_1^4c_2}{c_3^2c_4^2c_5}\,.
\eqlabel{c4def}
\end{equation}
From \eqref{effact} the inflaton potential is
\begin{equation}
V\equiv T_3\biggl(c_1^4-C_4\biggr)\,,
\eqlabel{pot}
\end{equation}
and the canonically normalized inflaton field $\Phi$ is given by
\begin{equation}
d\Phi=\sqrt{T_3}\ c_1c_2\ dr\,.
\eqlabel{cannom}
\end{equation}
Using UV solution of the de-Sitter deformed resolved warped deformed conifold of section 3.2 we find for $x\ll 1$
\begin{equation}
\begin{split}
\frac{d^2V}{d\Phi^2}=&2 \H^2-\frac{27 P^2 \H^4}{2\lambda^2}\ x^2\  \biggl(3 \ln x+4\biggr)+x^4\ \biggl(\frac{1701 \H^6 P^4}{40\lambda^4} \biggl(120 \ln^2 x+285 \ln x-31\biggr)\\
&-\frac{27 \H^4 P^2}{5\lambda^2} 
\biggl(40 \hw_{3,(2,1)} \ln^2 x+10 (2\hw_{3,(2,1)}- \ha_{(2,0)})\ln x-163 \ha_{(2,0)}-202 \hw_{3,(2,1)}\biggr)
\\
&+\H^2 \biggl(16 \hw_{3,(2,1)}^2 \ln^2 x+8(4 \hw_{3,(2,1)}^2- \hw_{3,(2,1)} \ha_{(2,0)} )\ln x-\frac{128}{5} \hh_{1,(4,0)}+\frac 45 \hw_{3,(2,1)}^2
\\
&-\frac{96}{5} g_{s,(4,0)}-\frac{136}{5} \ha_{(2,0)} \hw_{3,(2,1)}-\frac{43}{5} \ha_{(2,0)}^2\biggr)\biggr)
+\calo\left(x^6\ln^5 x\right)\,.
\end{split}
\eqlabel{final}
\end{equation}
Notice that to order $\calo(x^4\ln^2 x)$ two more UV parameters appear: $g_{s,(4,0)}$ and $h_{1,(4,0)}$.
Also,  $\frac{dV^2}{d\Phi^2}$ vanishes to this order whenever $\H=0$. If fact, we find that for the supersymmetric 
background of \cite{grana}
\begin{equation}
\frac{d^2V}{d\Phi^2}\bigg|_{susy}= \frac{\lambda^2\ha_{(2,0)}^4}{128P^2}\ x^6\ \biggl(8+15\ln x\biggr)+\calo(x^8\ln x)\,,
\eqlabel{grana}
\end{equation}
in agreement with the computation of \cite{dks}.

From \eqref{final} we conclude that inflation deep inside the warped throat of the resolved deformed conifold is fast-roll. The 
corresponding cosmological parameter $\eta$ is 
\begin{equation}
\eta=\frac 23+\calo\left(\Phi^{-2}\ln\Phi\right)\,.
\eqlabel{etaf}
\end{equation}
Result \eqref{etaf} is expected from the general arguments 
presented in \cite{bi0}.

\section{Conclusion}
In this paper we analyzed background curvature corrections in DKS inflationary model \cite{dks}.
We found that due to the conformal coupling to the background, the inflaton  in this model receives Hubble
scale mass correction which renders slow-roll inflation impossible. The correct slow-roll parameter $\eta$ 
in DKS model is given by \eqref{etaf}. While DKS inflationary scenario in strongly warped region of the 
geometry does not allow for tuning of $\eta$, the other slow-roll parameter\footnote{Given 
\eqref{final} it can be readily evaluated.} $\epsilon$,
\begin{equation}
\epsilon\equiv \frac{1}{18 M_{pl}^2}\ \left(\frac{1}{\H^2}\frac{dV}{d\Phi}\right)^2=\frac{2}{9}\frac{\Phi^2}{M_{pl}^2}+
\calo\left(\frac{\H^4}{M_{pl}^2\Phi^2}\ln^2\Phi\right)\,,
\eqlabel{ep}  
\end{equation}
can be made small. Indeed, one can always choose a baryonic branch deformation parameter $U$ of \cite{grana} such that
\begin{equation}
|U|\ll M_{pl}^2\,,
\eqlabel{uchoice}
\end{equation}
which would allow inflaton to be simultaneously in the highly warped region of the geometry $\Phi^2\gg |U|$
while having small second slow-roll parameter \eqref{ep}.

The fact 
that conformal coupling of the inflationary 3-brane to the background leads to unacceptably large $\eta$-parameter 
in generic supergravity backgrounds was emphasized in  \cite{bi0}. The obvious solution to this 
large-$\eta$ problem proposed in \cite{br} was 
to turn on appropriate 3-form fluxes to compensate for the 3-brane conformal coupling.
Explicit realization of the latter proposal was discussed in \cite{bg}, where the authors study 
inflation in de-Sitter deformed $\caln=2^*$ warped throat.

\section*{Acknowledgments}
It is a pleasure to thank Ofer Aharony, Micha Berkooz and Lev Kofman for stimulating discussions.
I would like to thank the Weizmann Institute of Science for hospitality during part of this work.
Research at Perimeter Institute is supported in part by funds from NSERC of
Canada. I gratefully   acknowledge  support by  NSERC Discovery
grant.

\end{document}